\documentclass{Interspeech}

\interspeechcameraready

\title{Improving Respiratory Sound Classification with Architecture-Agnostic Knowledge Distillation from Ensembles}

\author[affiliation={1}]{Miika}{Toikkanen}
\author[affiliation={1,2}]{June-Woo}{Kim$^\dagger$}
\affiliation{RSC LAB}{MODULABS}{Republic of Korea}
\affiliation{Department of Psychiatry}{Wonkwang University Hospital}{Republic of Korea}

\email{miika.toikkanen.2@gmail.com, kaen2891@gmail.com}
\usepackage{comment}
\usepackage{cite}
\usepackage{rotating}
\usepackage[table]{xcolor}
\usepackage{makecell}
\usepackage{tabu}
\usepackage{subcaption}
\usepackage{multicol}
\usepackage{multirow}

\begin{document}

\maketitle
\renewcommand{\thefootnote}{$\dagger$}
\footnotetext{Corresponding author.}
\renewcommand{\thefootnote}{\arabic{footnote}}

\begin{abstract}

    Respiratory sound datasets are limited in size and quality, making high performance difficult to achieve. Ensemble models help but inevitably increase compute cost at inference time. Soft label training distills knowledge efficiently with extra cost only at training. In this study, we explore soft labels for respiratory sound classification as an architecture-agnostic approach to distill an ensemble of teacher models into a student model. We examine different variations of our approach and find that even a single teacher, identical to the student, considerably improves performance beyond its own capability, with optimal gains achieved using only a few teachers. We achieve the new state-of-the-art Score of 64.39 on ICHBI, surpassing the previous best by 0.85 and improving average Scores across architectures by more than 1.16. Our results highlight the effectiveness of knowledge distillation with soft labels for respiratory sound classification, regardless of size or architecture.

    %

    %I think last sentence is good for emphasizing our work maybe? you think not helpful? In objective perspectives, yes, but it can be emphasized.
    % if you want we can add, i need to condense more to fit it
    %These results highlight the effectiveness of knowledge distillation with soft labels for respiratory sound classification, regardless of the size or architecture of the model.

\end{abstract}

\section{Introduction}

Respiratory sound classification (RSC) has been an active research area due to its potential to aid in diagnosing respiratory diseases. Previous works primarily focused on CNN architectures~\cite{yang2020adventitious, ma2020lungrn+, Gairola21, chang22h_interspeech, ren2022prototype, wang2022domain, nguyen2022lung, moummad2023pretraining}, such as ResNet~\cite{he2016deep}, EfficientNet~\cite{tan2019efficientnet}, and CNN6~\cite{kong2020panns}. More recently, pretrained Audio Spectrogram Transformer (AST)~\cite{gong21b_interspeech, bae23b_interspeech} models trained on large-scale datasets like ImageNet~\cite{deng2009imagenet} and AudioSet~\cite{audioset} have demonstrated the advantages of self-attention-based mechanisms. Building on AST, techniques such as Patch-Mix contrastive learning~\cite{bae23b_interspeech}, adversarial methods for synthetic sample inconsistencies~\cite{kim2023adversarial}, stethoscope bias mitigation~\cite{kim2024stethoscope}, and large-scale pretraining method~\cite{niizumi2024masked, zhang2024towards} have further advanced the field. Bridging the text and audio modalities (BTS)~\cite{kim24f_interspeech} to leverage textual metadata prompts has also shown notably boosted RSC performance, achieving a state-of-the-art score of 63.54\% on the ICBHI dataset~\cite{rocha2018alpha}.

However, high-quality open-source respiratory sound datasets are a scarce resource
%and often limited qualit
, which makes it difficult to train strong models. Ensemble models offer a simple way of boosting model performance, but at the cost of increased computation at inference time~\cite{li2023towards}. Knowledge distillation~\cite{hinton2015distilling, gou2021knowledge, wang2021knowledge} is a widely adopted technique for compressing deep neural network ensembles and transferring knowledge from complex or large-scale teacher models to more compact student models, thereby boosting their performance. One of the key mechanisms in knowledge distillation is employing the soft labels~\cite{zhou2021rethinking, yang2023knowledge, busbridge2025distillation}, which represent the probabilistic outputs of the teacher model. These soft labels provide richer representation than hard labels, enabling the student model to learn better representations of the data~\cite{zhou2021rethinking, yang2023knowledge}. Despite the success of distillation in various domains, its application to the RSC task remains relatively underexplored. 
We present a systematic evaluation of soft-label distillation in RSC, covering multiple ensemble configurations, and scales of models.
%We conduct the first systematic study of soft-label distillation in this domain, evaluating multiple ensemble configurations, training strategies, and model scales. Our findings offer practical insights into how to improve performance without increasing inference cost.

In this study, we address this gap by investigating the effectiveness of soft label distillation in the context of RSC as an architecture-agnostic approach to distill ensemble knowledge into individual models. 
% We thoroughly explore several configurations of teacher ensembles and their application to a range of model architectures.
%Our results show that even a single teacher can enhance classification performance beyond its own capability, with optimal gains achieved using only a few teachers and the large potential of a large well selected teacher ensemble. This method establishes a new state-of-the-art on the ICBHI benchmark, highlighting the potential of knowledge distillation and ensembling when working with noisy and small respiratory sound classification datasets.
Our main contributions are summarized as follows:
\begin{itemize}
% even stronger model from distilled!
\item We demonstrate the strength of soft label distillation from ensembles on RSC data and set the new SOTA score 64.39 on the ICBHI dataset.
\item We explore methods for creating the soft labels and teacher ensembles. We find that even a single teacher is able to boost the performance of identical students, and just a few are enough for optimal performance, with additional benefits gained by curating the ensemble.
\item We also investigate compressing a second-generation ensemble from a first-generation teacher ensemble and find it to produce considerably stronger predictor, increasing the ICBHI score from 64.34 to 65.45.
\item We release our code to support reproducibility and further research at \\\url{https://github.com/RSC-Toolkit/rsc-ensemble-kd}.

\end{itemize}
\section{Preliminaries}
\subsection{Dataset Description}
The ICBHI respiratory sound dataset~\cite{rocha2018alpha} is a widely recognized benchmark for RSC tasks. It consists of approximately 5.5 hours of respiratory sound recordings, comprising a total of 6,898 breathing cycles. The dataset is officially divided into training (60\%) and testing (40\%) subsets at the breathing cycle level, ensuring no patient overlap between splits. Specifically, the training set contains 4,142 cycles, while the test set includes 2,756 cycles, categorized into four distinct classes: \textit{normal}, \textit{crackle}, \textit{wheeze}, and \textit{both} (crackle and wheeze). 
We binarized the age groups into adults (over 18 years old) and pediatrics (18 years old or younger), following the approach in previous studies~\cite{kim24f_interspeech}. Other metadata attributes, including sex, recording location, and recording device, are maintained as per the original ICBHI annotations. 

\subsection{Training Details}
We applied data pre-processing to extract the respiratory cycles from the waveform samples, ensuring each cycle was standardized to 8 seconds, as described in~\cite{bae23b_interspeech, kim2023adversarial, 10782363, kim24f_interspeech, Gairola21}. All samples were resampled to 16 kHz, except when employing the BTS model, which operates at 48 kHz. For the BTS model experiments, we adopted the same experimental settings as described in~\cite{kim24f_interspeech}. The Adam optimizer was employed with a learning rate of 5e--5, cosine scheduling, and a batch size of 8 for fine-tuning the transformer based models (BTS, CLAP~\cite{wu2023large} and AST) until 50 epochs. For other architectures, a larger learning rate of 1e--3 and a batch size of 128 was used, with training conducted for 200 epochs. SpecAugment~\cite{park2019specaugment} is applied to all the architectures except for BTS and CLAP.

\subsection{Evaluation Metrics}
We employ \textit{Sensitivity} ($S_e$), \textit{Specificity} ($S_p$), and their arithmetic mean, referred to as ICBHI \textit{Score} metrics, by the standard definitions provided in~\cite{rocha2018alpha}. $S_e$ and $S_p$ represent the proportion of actual respiratory abnormality and healthy cases that are correctly classified, respectively. We report both the mean and variance of $S_p$, $S_e$, and Score across multiple independent runs, each initialized with different seeds $\{1, 2, 3, 4, 5\}$. In case of ensembles, the metrics are reported from single run.
\section{Method}

\begin{figure}[t!]
    \centering
    \includegraphics[width=1.0\linewidth]{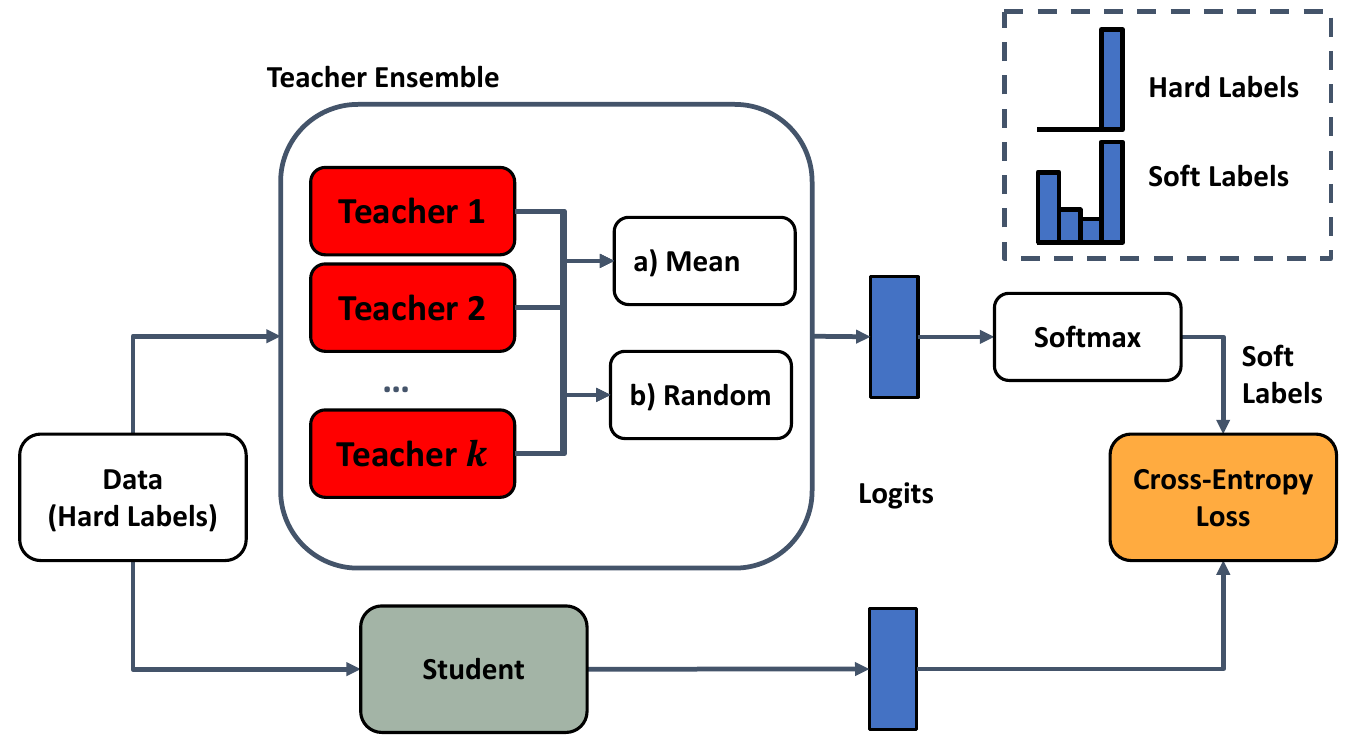}
    \caption{Diagram of the overall method. We create soft labels a) by taking the mean, or b) sampling the label from random predictor. The soft label replaces the hard label as training target for cross-entropy loss. Hard labels exist only in the source data and are not used during the process.}
    \label{fig:architecture}
    \vspace{-3mm}
\end{figure}

We employ a response-based knowledge distillation approach, where the teacher is an ensemble of predictors as depicted in Figure \ref{fig:architecture}, and the student model learns to mimic the teacher's responses.
%Soft labels represent probabilities of the target classes and can capture similarity of input to several classes as opposed to the hard label that are a one-hot vector. 
The teacher ensemble creates soft labels by computing the mean of the logits from all predictors, or by sampling the logits of a single predictor from the ensemble. The following sections explain the process in detail.

\subsection{BTS Model}
The BTS (Bridging Text and Sound)~\cite{kim24f_interspeech} model is a multimodal framework designed to integrate respiratory sound with textual metadata prompts to boost respiratory sound classification performance. Derived from the pretrained LAION-CLAP~\cite{wu2023large} model, BTS employs an acoustic encoder to extract robust respiratory sound features, while its text encoder processes metadata as textual prompts with details, such as patient age groups, sex, recording stethoscopes, and recording locations. By aligning these modalities within a shared latent space, 
%in a multimodal manner
%BTS has showcased state-of-the-art RSC performance.
%on the ICBHI dataset. 
%Therefore, we adopt BTS as a strong teacher model for knowledge distillation, leveraging its rich multimodal feature representations to improve the performance of distilled or lightweight student models. 
This multimodal approach achieves state-of-the-art results on the ICBHI dataset, but has a higher compute cost in comparison to the other RSC models, making it a good choice as a teacher for knowledge distillation.

\subsection{BTS++: Ensemble of Strong Models}

% can condense like this
We train 30 BTS models with different seeds. The mean ICBHI score of all 30 individual models of the ensemble is 63.41 $\pm \text{0.77}$.
At test time, ensemble predictions are computed by averaging logits and selecting the class corresponding to the averaged logit with the highest magnitude.
For each sample in the data, we denote the logits as \(z \in \mathbb{R}^{N \times C}\), where \(N\) is the number of models and \(C\) the number of classes and $i$-th logits in $z$ are represented as $z_i$.
We use the suffix ``++" to denote the ensembled version of a model and $k$ to denote the number of predictors, so the BTS ensemble with 5 predictors is denoted as BTS++[${k=5}$].
%We train BTS models with 30 different seeds and compute their logits for each sample of the training dataset. The logits \(z_{i} \in \mathbb{R}^{C}\) correspond to the $i$th model in the ensemble and the full set of logits for each sample is represented as \(z \in \mathbb{R}^{N \times C}\), where \(N\) is the number of models, and \(C\) is the number of classes. We use $k$ to denote the number of teachers selected from all possible teachers $\{1, \dots, N\}$. The ensembles built from these logits are denoted with BTS++[k], where $k$ is the number of models.
% mean, std, max, min, 63.41233333	0.7793750075	64.91	61.89
%The mean ICBHI score of the all individual models in the ensemble is 63.41 (+/- 0.77) and the models are trained using the original BTS training code, with the seeds ranging from 1 to 30 in order, such that for $k$, we select seeds from 1 to $k$.
%At test time, we combine the ensemble labels by taking the mean of the ensemble logits, and selecting the index of the highest magnitude as the class index. 

\subsection{BTS-d: Soft Label Distillation}

Following response-based knowledge distillation principles, we generate soft labels \(p \in \mathbb{R}^{C}\) from teacher logits $z$ using two approaches, \emph{mean teacher} and \emph{random teacher}.
We construct the labels separately for each iteration, such that random sampling results are different for each epoch
%$$, which creates a mean label, and \emph{random teacher}, which creates a slightly noisier random label by sampling. 

\textbf{Mean Teacher}. In this case, we choose all models in range $\{1, \dots, k\}$, compute the mean logits and apply the softmax function to obtain $p_{\mu}$. This represents the mean opinion of all $k$ possible teachers as a probability distribution.
\begin{equation}
p_{\mu} = \text{softmax} \left( \frac{1}{k} \sum_{i=1}^{k} z_i \right)
\end{equation}

\textbf{Random Teacher}. In this case we uniformly sample a teacher index \(i\) from \(\{1, \dots, k\}\) and apply the softmax function to obtain the random teacher label \(p_r\). This label represents the probability distribution corresponding to one of the \(k\) possible teachers.
\begin{equation}
    p_r = \text{softmax}(z_i), \quad i \sim \text{Uniform}(\{1, \dots, k\})
\end{equation}
%\begin{align}
    %i &\sim \text{Uniform}(\{1, \dots, k\}) \\
    %p_r &= \text{softmax}(z_i)
%\end{align}

\textbf{Knowledge Distillation.} 
Given the student model predictions \(\hat{y}\), we optimize its parameters using the cross-entropy loss \(\mathcal{L}_{\text{CE}}\) with soft labels from the teacher model instead of the hard labels \(y\). In this process, the hard labels are not used at all. The resulting knowledge distillation losses are $\mathcal{L_{\mu}}$ for the mean teacher, and $\mathcal{L_{\mathrm{r}}}$ for the random teacher.
%\(\mathcal{L}_{\text{S}}\), where \(p\) represents either \(p_{r}\) (random teacher) or \(p_{\mu}\) (mean teacher), guides the student model’s learning, enabling effective knowledge transfer from the teacher model without an additional hard-label loss term.
%Given the student model predictions $\hat{y}$, we optimize its parameters using the Cross-Entropy loss $\mathcal{L_{\text{CE}}}$, but replace the hard label $y$ with the soft labels. We do not use the hard label $y$ at all during the training of the student model. The resulting knowledge distillation loss $\mathcal{L_\text{S}}$, where p is one of  optimizes $p_r$ or $p_{\mu}$, optimizes the student model parameters, ensuring the effective transfer of knowledge from the teacher model without an additional loss term for hard labels.
%while enhancing the student's learning efficiency.
\begin{equation}
    \mathcal{L_{\text{CE}}} = H(y, \hat{y}) = -\sum_{i=1}^n \, y_{i}\! \, \log \, (\hat{y}_{i})
\end{equation}
\begin{equation}
    \mathcal{L_{\mu}} = H(p_{\mu}, \hat{y})
\end{equation}
\begin{equation}
    \mathcal{L_{\mathrm{r}}} = H(p_r, \hat{y})
\end{equation}

We use the suffix ``-d" to denote the distilled version of a model and $k$ to denote the number of teachers, so the BTS student with 5 teachers is denoted as BTS-d[${k=5}$].

%Softmax on teacher logits is essential for stable training, as models trained on raw logits fail to converge as we find in the ablation study. This aligns with findings on response-based knowledge distillation, where soft targets are a probability distribution.
%Applying the softmax operation to the teacher model logits is necessary to stabilize training, we found that the models trained with raw teacher logits failed to converge under the cross-entropy learning objective, even when using the mean teacher labels, which will be demonstrated in our Experiments Section. 

%\subsection{Lightweight Distillation}
%We extend soft label distillation to lightweight architectures, where the architecture of the student is not the same as the architecture of the teacher, enabling performance improvements on smaller models while retaining efficient inference.

\subsection{BTS-d++: Ensemble of Even Stronger Models}

To further improve ensemble model performance, we create a second-generation ensemble by combining the BTS-d models distilled from the BTS++ teacher. This compresses the teacher ensemble into a student ensemble BTS-d++, that comprises of even stronger individual predictors.

\begin{table*}[!t]
    \centering
    %\vspace{-3mm}
    \caption{Main results comparing models on the ICBHI dataset with the official train-test split. 
    Pretraining Data column, IN, AS, and LA refer to ImageNet~\cite{deng2009imagenet}, AudioSet~\cite{audioset}, and LAION-Audio-630K~\cite{wu2023large}, respectively. $*$ denotes the previous state-of-the-art ICBHI Score.
    The \textbf{Best} and {\underline{second best}} non-ensemble results are highlighted by the bold characters and underlines.
    The suffixes "-d" and "++" indicate distillation and ensembling, respectively.
    }
    
    \label{table1_main_results}
    \renewcommand{\arraystretch}{1}
    \addtolength{\tabcolsep}{8pt}
    \resizebox{\linewidth}{!}{
    \begin{tabular}{p{2pt}llll|lll}
    \toprule
    & Method & Backbone & Pretraining Data & Venue & $S_p$\,(\%) & $S_e$\,(\%) & \textbf{Score}\,(\%) \\
    \hline \midrule

    \multirow{18.5}{*}{\rotatebox[origin=c]{90}{\textbf{4-class eval.}}} & SE+SA \cite{yang2020adventitious} & ResNet18 & - & \textit{INTERSPEECH`20} & {81.25} & 17.84 & 49.55 \\

    & LungRN+NL \cite{ma2020lungrn+} & ResNet-NL & - & \textit{INTERSPEECH`20} & 63.20 & 41.32 & 52.26 \\
    
    & RespireNet \cite{Gairola21} (CBA+BRC+FT) & ResNet34 & IN & \textit{EMBC`21} & 72.30 & 40.10  & 56.20 \\

    & Chang \textit{et al.} \cite{chang22h_interspeech} & CNN8-dilated & - & \textit{INTERSPEECH`22} & 69.92 & 35.85 & 52.89 \\
    
    & Ren \textit{et al.} \cite{ren2022prototype} & CNN8-Pt & - & \textit{ICASSP`22} & 72.96 & 27.78 & 50.37 \\
    
    & Wang \textit{et al.} \cite{wang2022domain} (Splice) & ResNeSt & IN & \textit{ICASSP`22} & 70.40 & 40.20 & 55.30 \\

    & Nguyen \textit{et al.} \cite{nguyen2022lung}\,(StochNorm) & ResNet50 & IN & \textit{TBME`22} & 78.86 & 36.40 & 57.63 \\
    
    & Nguyen \textit{et al.} \cite{nguyen2022lung}\,(CoTuning) & ResNet50 & IN & \textit{TBME`22} & 79.34 & 37.24 & $\text{58.29}$ \\
   
    & Moummad \textit{et al.} \cite{moummad2023pretraining}\,(SCL) & CNN6 & AS & \textit{WASPAA`23} & 75.95 & 39.15 & 57.55 \\    
    
    & Bae \textit{et al.} \cite{bae23b_interspeech}\, (Fine-tuning)  & AST & IN\,+\,AS & \textit{INTERSPEECH`23} & $\text{77.14}$ & $\text{41.97}$ & $\text{59.55}$ \\
    
    & Bae \textit{et al.} \cite{bae23b_interspeech}\, (Patch-Mix CL) & AST & IN\,+\,AS & \textit{INTERSPEECH`23} & $\text{81.66}$ & $\text{{43.07}}$ & $\text{62.37}$ \\
    
    & Kim \textit{et al.} \cite{kim2023adversarial}\, (AFT on Mixed-500) & AST & IN\,+\,AS & \textit{NeurIPSW`23} & $\text{{80.72}}$ & $\text{{42.86}}$ & $\text{61.79}$ \\
    
    & Kim \textit{et al.} \cite{kim2024stethoscope}\, (SG-SCL) & AST & IN\,+\,AS & \textit{ICASSP`24} & $\text{{79.87}}$ & $\text{{43.55}}$ & $\text{{61.71}}$ \\

    & Kim \textit{et al.} \cite{10782363}\, (RepAugment) & AST & IN\,+\,AS & \textit{EMBC`24} & $\text{82.47}$ & $\text{40.55}$ & $\text{61.51}$ \\

    & Daisuke \textit{et al.} \cite{niizumi2024masked}\, (M2D-X/0.7) & M2D ViT & AS & \textit{TASLP`24} & $\text{81.51}$ & $\text{{45.08}}$ & $\text{63.29}$ \\

    & Kim \textit{et al.} \cite{kim24f_interspeech}\, (Audio-CLAP) & CLAP & LA & \textit{INTERSPEECH`24} & $\text{80.85}$ & $\text{44.67}$ & $\text{62.56}$ \\

    & Kim \textit{et al.} \cite{kim24f_interspeech}\, (BTS) & CLAP & LA & \textit{INTERSPEECH`24} & $\text{81.40}$ & $\text{\underline{45.67}}$ & $\text{63.54}^\textbf{*}$ \\

    %\midrule

    %\multicolumn{7}{c}{Distilled Models [ours]} \\
    
    \midrule

    \multirow{4}{*}{\rotatebox[origin=c]{90}{\textbf{Distill.}}} & \textbf{AST-d[${k=5}$] (mean teacher) [ours]} & \text{AST} & IN\,+\,AS & - & $\text{79.24}_{\pm 3.51}$ & 
    $\text{42.92}_{\pm 4.04}$ & 
    $\text{61.08}_{\pm 1.26}$ \\

    & \textbf{Audio-CLAP-d[${k=5}$] (mean teacher) [ours]} & \text{CLAP} & LA & - & $\text{82.82}_{\pm 2.14}$ & 
    $\text{44.44}_{\pm 2.35}$ & 
    $\text{63.63}_{\pm 0.60}$ \\
    
    & \textbf{BTS-d[${k=5}$] (mean teacher) [ours]} & \text{CLAP} & LA & - & $\textbf{84.93}_{\pm 2.25}$ & 
    $\text{43.82}_{\pm 2.25}$ & 
    $\text{\underline{64.38}}_{\pm 0.36}$ \\   
    
    & \textbf{BTS-d[${k=15}$] (random teacher)} [ours] & \text{CLAP} & LA & - & $\text{\underline{82.89}}_{\pm 2.14}$ & 
    $\textbf{45.90}_{\pm 1.89}$ & 
    $\textbf{64.39}_{\pm 0.42}$ \\

    %\midrule

    %\multicolumn{7}{c}{Ensemble Models [ours]} \\
    
    \midrule

    \multirow{4}{*}{\rotatebox[origin=c]{90}{\textbf{Ensemble.}}} & \textbf{BTS++[${k=5}$] [ours]} & \text{CLAP} & LA & - & $\text{85.18}$ & 
    $\text{43.50}$ & 
    $\text{64.34}$ \\

    & \textbf{BTS-d++[${k=5}$] (Second Generation) [ours]} & \text{CLAP} & LA & - & $\text{88.09}$ & 
    $\text{42.82}$ & 
    $\text{65.45}$ \\

    & \textbf{BTS++[${k=15}$] [ours]} & \text{CLAP} & LA & - & $\text{88.28}$ & 
    $\text{41.21}$ & 
    $\text{64.75}$ \\

    & \textbf{BTS++[${k=30}$] [ours]} & \text{CLAP} & LA & - & $\text{89.49}$ & 
    $\text{41.89}$ & 
    $\text{65.69}$ \\

%    \midrule

    \bottomrule
    \end{tabular}}
    %\vspace{-5mm}
    
\end{table*}
\section{Experiments}

%\subsection{Comparison with Previous Work}
% add both 5 mean and 15 random. Explain the easoning because ablation was best
% add 5, 15, 30 ensembles?
\subsection{Main Results}
Table \ref{table1_main_results} presents the main results compared to previous RSC work on the ICBHI dataset. 
For a fair comparison, we separate the ensemble models (labeled with ``++") from the non-ensemble models as they multiply the test-time compute cost.
Without additional test-time compute cost, the distilled BTS-d outperforms all previous approaches. Highest score was achieved with random teacher ${k=15}$, increasing the previous state-of-the-art score by 0.85 from 63.54 to 64.39.
The mean teacher with ${k=5}$ is nearly as strong in terms of score, reaching 64.38.
Note that the optimal $k$ values for each teacher method were chosen based on empirical results in Figure \ref{fig:student_score}.
Considering the ensembles, BTS++[${k=5}$] reaches the ICBHI Score of 64.34, but also multiplies the test-time compute cost by 5. 
The second generation ensemble BTS-d++[${k=5}$] reaches 65.45 with the same compute cost, highlighting the value of good quality teachers.
Increasing $k$ to 15 increases the score to 64.75 and further increasing $k$ to 30 for the full available ensemble raises the score to 65.69, but with a massive 30x compute cost over the baseline BTS.
Based on these results, for practical real-time inference, our soft-label distilled results BTS-d and the second generation ensemble BTS-d++ strike a good balance between performance and inference cost.

\subsection{Effectiveness of Distillation on Lightweight Architectures}
% lightweight
As the method is architecture-agnostic, we investigate the effectiveness of soft-label distillation on various architectures, some of which are much smaller than the teacher models.
We select some of the models commonly used for RSC in previous work~\cite{kim24f_interspeech, 10782363} and apply soft label training with BTS++[${k=5}$] as the mean teacher.
Table \ref{table2_soft_hard_comp} compares standard cross-entropy with hard label (one-hot) against the soft label distillation on ICBHI.
Each of the models, regardless of architecture or size benefit from soft-label distillation. 
While on average the score increased by 1.16, generally the specificity increased much more, and the sensitivity decreased slightly, indicating a trade-off between reducing false positives and a slight increase in false negatives. This suggests that soft-label distillation helps the models better distinguish normal from abnormal cases.
From these results we also see that, compact models with lower inference cost, such as ResNet18~\cite{he2016deep}, EfficientNet~\cite{tan2019efficientnet}, CNN6~\cite{kong2020panns}, and Audio-CLAP~\cite{kim24f_interspeech, wu2023large} can approach the performance of larger, more computationally expensive teacher model. 
%...

\begin{table*}[!t]
    \centering
    %\vspace{-3mm}
    \caption{Comparison of hard and soft labels for different architectures on the ICBHI dataset for the official 60--40\% train--test split task of respiratory sound classification.}% $*$ denotes our implementation because the BTS model has randomness results as described in~\cite{kim24f_interspeech}.}
    
    %\label{table1_main_results}
    \label{table2_soft_hard_comp}
    \renewcommand{\arraystretch}{1}
    \addtolength{\tabcolsep}{8pt}
    \resizebox{\linewidth}{!}{
    \begin{tabular}{ll|ccc|ccc|c}
    \toprule
    Model & \# Params & \multicolumn{3}{c}{Hard Label} & \multicolumn{3}{c}{Soft Label} & Gain \\
    & & $S_p$\,(\%) & $S_e$\,(\%) & \textbf{Score}\,(\%) & $S_p$\,(\%) & $S_e$\,(\%) & \textbf{Score}\,(\%) & \textbf{Score}\,(\%) \\
    \hline \midrule

    ResNet18  & 11.7M  & $\text{76.70}_{\pm 5.65}$ & $33.47_{\pm 4.03}$ & $\text{55.09}_{\pm 0.82}$ & $\text{81.41}_{\pm 2.78}$ & $\text{31.11}_{\pm 3.05}$ & $\text{56.26}_{\pm 0.85}$ & \text{1.17}  \\
    EfficientNet & 5.3M & $\text{78.21}_{\pm 3.56}$ & $\text{34.44}_{\pm 2.84}$ & $\text{56.33}_{\pm 0.43}$ & $\text{79.26}_{\pm 3.31}$ & $\text{36.09}_{\pm 3.39}$ & $\text{57.68}_{\pm 1.48}$ & \text{1.35}  \\
    CNN6 & 4.8M & $\text{77.00}_{\pm 3.27}$ & $\text{37.35}_{\pm 3.15}$ & $\text{57.17}_{\pm 0.81}$ & $\text{85.57}_{\pm 2.56}$ & $\text{30.77}_{\pm 1.97}$ & $\text{58.17}_{\pm 0.60}$ & \text{1.00}  \\

    AST & 87.7M        & $\text{77.14}_{\pm 3.35}$ & $\text{43.07}_{\pm 2.80}$ & $\text{59.55}_{\pm 0.88}$ & $\text{79.24}_{\pm 3.51}$ & $\text{42.92}_{\pm 4.04}$ & $\text{61.08}_{\pm 1.26}$ & \text{1.53}  \\
    Audio-CLAP & 28M   & $\text{80.85}_{\pm 3.33}$ & $\text{44.67}_{\pm 3.77}$ & $\text{62.56}_{\pm 0.37}$ & $\text{82.82}_{\pm 2.14}$ & $\text{44.44}_{\pm 2.35}$ & $\text{63.63}_{\pm 0.60}$ & \text{1.07}  \\
    BTS$^{*}$ & 153M        & $\text{81.40}_{\pm 2.57}$ & $\text{45.67}_{\pm 2.66}$ & $\text{63.54}_{\pm 0.80}$ & $\text{82.89}_{\pm 2.14}$ & $\text{45.90}_{\pm 1.89}$ & $\text{64.39}_{\pm 0.42}$ & \text{0.85}  \\

    \midrule

    Average & - & $\text{78.55}_{\pm 3.62}$ & $\text{39.78}_{\pm 3.21}$ & $\text{59.04}_{\pm 0.69}$ & $\text{81.87}_{\pm 2.74}$ & $\text{38.54}_{\pm 2.78}$ & $\text{60.20}_{\pm 0.87}$ & 1.16 \\
    
    \bottomrule
    \end{tabular}}
    %\vspace{-5mm}
    
\end{table*}

\subsection{Ablation Study}

We perform an ablation study to verify the method and report the results in Table \ref{table3_ablation}. 
First, we compare teacher labels against random noise.
Noised Label ($\text{var}=0.1$) adds a Gaussian noise vector with variance of 0.1 across training target label.
Noised label (teacher var) models the variance of mean teacher logits on the training dataset and adds the equivalent noise vector to the training target label.
The noised labels perform worse than the baseline, indicating that the softness of the labels is not the reason for increased performance, but modeling the teacher variance does shift the sensitivity closer to the teacher's sensitivity value.
Using a single teacher to create the soft labels boosts the score notably (63.54\% to 63.90\%), and further using a teacher ensemble yields the best performance, 64.38\% and 64.39\% for mean teacher at ${k=5}$ and random teacher at ${k=15}$, respectively.
%indicating more useful information gets added to the labels.
Selecting the best models as teachers is also beneficial. We picked the 5 highest score checkpoints as the curated mean teacher ensemble and further raised the score to 64.61\%.
Since the soft label is only applied during training, this can be considered a valid approach and does not cause leakage of test data.
We also remove the softmax from the soft labels and instead train using the raw teacher ensemble logits as labels, 
but the model fails to converge properly with cross-entropy loss.
%but this prevents the model from converging well with cross-entropy.
These results clearly indicate that the teacher ensemble is useful for creating the soft labels as formulated in this work.

\begin{table}[!t]
    \centering
    %\vspace{-3mm}
    \caption{Ablation study of BTS-d performance comparing different variations of the soft labels.}
    
    %\label{table1_main_results}
    \label{table3_ablation}
    \renewcommand{\arraystretch}{1}
    \addtolength{\tabcolsep}{1pt}
    \resizebox{\linewidth}{!}{
    \begin{tabular}{l|lll}
    \toprule
    Method           & $S_p$\,(\%) & $S_e$\,(\%) & \textbf{Score}\,(\%)        \\
    \hline \midrule
    Baseline         & $\text{81.40}_{\pm 2.57}$ & $\text{45.67}_{\pm 2.66}$ & $\text{63.54}_{\pm 0.80}$ \\
    Noised Label (var=0.1)     & $\text{81.33}_{\pm 2.90}$ & $\text{44.10}_{\pm 3.53}$ & $\text{62.71}_{\pm 0.47}$ \\    
    Noised Label (teacher var)     & $\text{76.53}_{\pm 3.78}$ & $\text{47.46}_{\pm 2.36}$ & $\text{62.00}_{\pm 1.06}$ \\
    Single Teacher      & $\text{83.18}_{\pm 1.59}$ & $\text{44.62}_{\pm 1.56}$ & $\text{63.90}_{\pm 0.15}$ \\
    Mean Teacher Ensemble (${k=5}$)   &  $\text{84.93}_{\pm 2.25}$ & $\text{43.82}_{\pm 2.25}$ & $\text{64.38}_{\pm 0.36}$ \\
    Random Teacher Ensemble (${k=15}$)    & $\text{82.89}_{\pm 2.14}$ & $\text{45.90}_{\pm 1.89}$ & $\text{64.39}_{\pm 0.42}$  \\
    Curated Teacher Ensemble (${k=5}$)    & $\text{84.28}_{\pm 2.58}$ & $\text{44.95}_{\pm 2.57}$ & $\text{64.61}_{\pm 0.75}$  \\
    Remove softmax    & 
    %\multicolumn{3}{c}{Failed to converge} \\
    %2nd Generation Student    & 
    \multicolumn{3}{c}{Failed to converge} \\
    \bottomrule
    \end{tabular}}
    %\vspace{-5mm}
    
\end{table}

\subsection{Diminishing Returns of Growing Teacher Ensemble Size}
\begin{figure}[t!]
    \centering
    \includegraphics[width=0.95\linewidth]{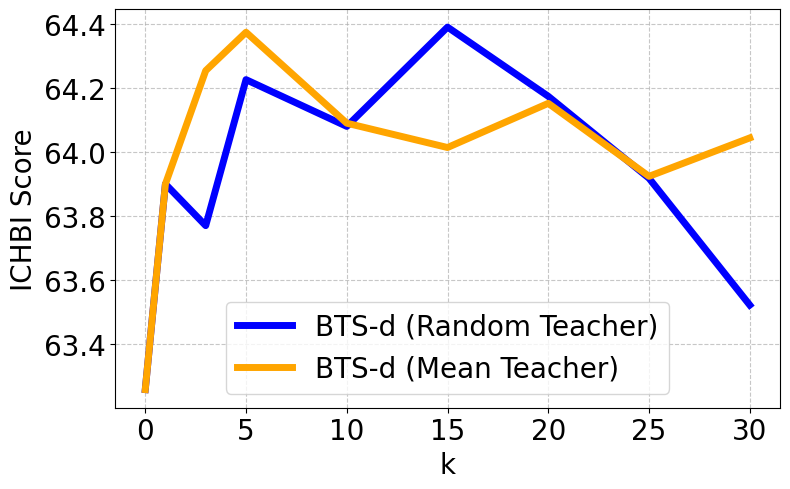} %\hspace{0.05\linewidth}
    
    \caption{The diminishing beneficial effect of increasing teacher count $k$.
    %Both the mean and random teacher methods are displayed.
    %, as well %as their mean smoothed with Gaussian filter (sigma=1).
    }
    \label{fig:student_score}
    \vspace{-3mm}
\end{figure}

In this section, we perform a series of experiments to study the optimal size for the teacher ensemble.
We train and evaluate a student model with $k$ set at different multiples of 5 from 0 teachers to 30 teachers, and include $k=1$ and $k=3$ to better view the effect at the lower end of the scale. The results are plotted in Figure \ref{fig:student_score}.
Each step is the mean Score of student model trained on 5 seeds (1-5).
Both the random and mean teacher methods benefit early and begin tapering off after their peaks. Mean teacher reaches the best score at 5 teachers, while random teacher peaks later at 15. 
%Both plots shows a decreasing trend after $k$ is increased beyond 10, while the test-time compute cost remains constant.
The random teacher reached a slightly higher value than mean teacher, but we have used the mean teacher with ${k=5}$ as the default soft label setting for our experiments because random teacher requires larger $k$ value to reach similar performance.

\begin{figure}[t!]
    \centering
    \includegraphics[width=1.0\linewidth]{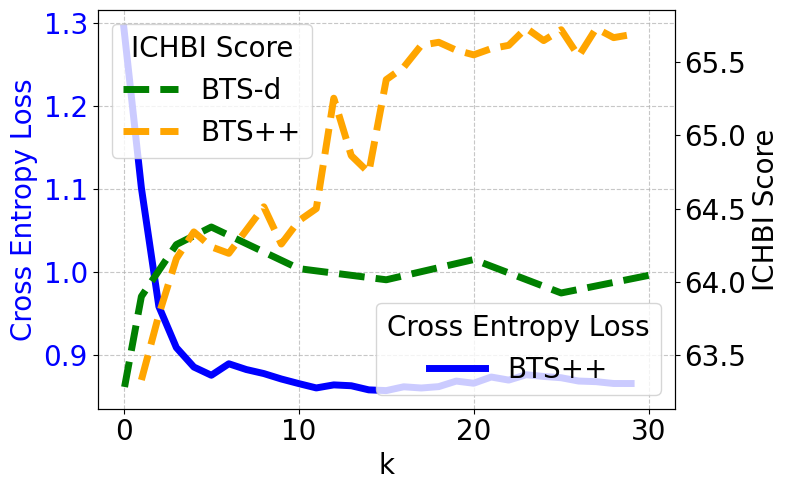}
    \caption{Teacher model BTS++ validation loss and ICBHI score compared to the student BTS-d ICBHI score as $k$ increases.}
    \label{fig:teacher_loss}
    \vspace{-3mm}
\end{figure}

We also evaluated the ensembled model BTS++ at each $k \in [1...30]$. Figure~\ref{fig:teacher_loss} compares the ICBHI score and validation loss of teacher ensemble BTS++, as well as the ICBHI score of distilled student model BTS-d plotted against increasing $k$. Growing the ensemble size yields increasing scores leveling off at around 65.7, while the test-time compute cost grows linearly as $k$ increases.
Although the score of the ensemble keeps increasing throughout the range of $k$, the score of student model tapers off after the teacher validation loss converges. At ${k=5}$, the distilled model performs similarly to the teacher ensemble, despite being only fifth of the size.
The reduced score could indicate that beyond this point, the student begins to overfit to teacher label distribution on training set, and cannot generalize more to the real ground truth distribution on test set. Therefore, under this setting, the optimal teacher ensemble is smaller in size than the full ensemble.
%..

\section{Conclusion}

In this study, we applied architecture-agnostic knowledge distillation for RSC using soft label training to extract knowledge from teacher ensembles. Our approach effectively transferred knowledge from the ensemble of teacher models to lightweight student models, achieving state-of-the-art performance on the ICBHI dataset. We demonstrated that even a single teacher model can significantly boost student model performance, with further gains observed when employing multiple teachers. We also found that curated 
%teacher ensembles
and second-generation ensembles further improved model performance. 
While we focus on RSC, the results are applicable to other classification tasks where data scarcity and inference constraints are critical, such as heart sound, ECG, or pathological speech classification.
%where lightweight deployment is needed due to real-time or hardware limitations.
Future research could explore diverse model architectures, as well as reducing the gap between ensemble and distilled model performance.
To encourage further research and support reproducibility, we release our code to the research community.

\section{Acknowledgement}
This research was supported by Brian Impact Foundation, a non-profit organization dedicated to the advancement of science and technology for all.

%%%%%

\bibliographystyle{IEEEtran}
\bibliography{mybib}

\end{document}